\documentclass[runningheads]{llncs}

\usepackage{graphicx}    
\usepackage{xcolor}      
\usepackage{subcaption}
\usepackage{algorithm}   
\usepackage{multicol}
\usepackage{algpseudocode}
\usepackage{multirow}
\usepackage[bottom]{footmisc} 

\usepackage[breaklinks,hidelinks]{hyperref} 

\begin{document}
\title{Enhancing Resource Management through Prediction-based Policies}
\titlerunning{Prediction-Based Policies for Resource Management}
\author{
    Antoni Navarro\inst{1}\orcidID{0000-0002-8317-6022} \and
    Arthur F. Lorenzon\inst{2}\orcidID{0000-0002-2412-3027} \and
    Eduard Ayguad\'e\inst{1}\orcidID{0000-0002-5146-103X} \and
    Vicen\c{c} Beltran\inst{1}\orcidID{0000-0002-3580-9630}
}
\authorrunning{A. Navarro et al.}
\institute{
    Barcelona Supercomputing Center, Barcelona, Spain \email{\{antoni.navarro,eduard.ayguade,vbeltran\}@bsc.es} \and
    Federal University of Pampa, Alegrete, RS, Brazil
    \email{aflorenzon@unipampa.edu.br}
}

\maketitle

\begin{abstract}
Task-based programming models are emerging as a promising alternative to make the most of multi-/many-core systems. These programming models rely on runtime systems, and their goal is to improve application performance by properly scheduling application tasks to cores. Additionally, these runtime systems offer policies to cope with application phases that lack in parallelism to fill all cores. However, these policies are usually static and favor either performance or energy efficiency. In this paper, we have extended a task-based runtime system with a lightweight monitoring and prediction infrastructure that dynamically predicts the optimal number of cores required for each application phase, thus improving both performance and energy efficiency. Through the execution of several benchmarks in multi-/many-core systems, we show that our prediction-based policies have competitive performance while improving energy efficiency when compared to state of the art policies.

\keywords{Energy efficiency \and Resource management \and Resource sharing \and OmpSs-2 \and Predictions \and Monitoring \and Cost}

\end{abstract}
\section{Introduction}
\label{section:introduction}

High-performance computing (HPC) systems are widely used to execute applications from many domains, such as financial computing, medical applications, and video and image processing. These systems are usually based on many-/multi-core architectures with heterogeneous memory and computing devices. Often, this implies the existence of complex memory hierarchies and technologies that evolve each year. Hence, application developers need productive and efficient tools to keep pace with the growing power of HPC systems. Task-based programming models have emerged as a promising alternative to develop complex applications on those systems. These models provide high-level abstractions to increase the productivity of application developers, and they rely on runtime systems to cope with system complexity. The main goal of a runtime system is to dynamically schedule application tasks to cores to optimize performance. However, these runtime systems must also cope with application phases with low parallelism that leave some of the cores without any task to execute. In this scenario, runtimes implement resource management policies to handle idle cores.

Commonly, resource managing policies focus on either improving performance or energy efficiency. Policies aiming to improve application performance adopt greedy strategies that always use all the available computational resources. A clear example would be OpenMP's~\cite{openmp} active policy, in which idle threads are actively checking for new work, consuming processor cycles and energy. On the other hand, techniques such as the ones in OpenMP's passive policy are used when the goal is to improve energy efficiency. In this case, idle threads immediately yield the processor to avoid contention inside the runtime and minimize the energy consumed. However, neither of these policies is adaptive enough to optimize both energy consumption and performance.

In order to tackle this challenge, two different hybrid approaches have been explored in the past~\cite{boguslavsky1994optimal,yan2016proposal}. The first one tries to improve performance-driven policies by adopting a greedy strategy for some time and, if no work is found in this period, yielding the processor to minimize energy consumption. Although it has positive effects on energy efficiency, the explored proposals struggle to find an optimal frequency to switch between policies. The second approach is based on policies that favor energy efficiency, in which idle resources are woken up at a specific frequency to check if new work is available. Similarly for both, finding a frequency that suffices all cases is a hassle.

In this work, we propose a novel resource management policy that can simultaneously optimize performance and energy efficiency. Our policy relies on the information provided by our monitoring and prediction framework to dynamically predict the number of cores that are required for each application phase. The main contributions of this work are: (\textit{i}) the creation of the monitoring and prediction infrastructure, which is capable of making precise workload predictions for task-based programming models; (\textit{ii}) the design of prediction-based resource managing policies; and (\textit{iii}) the enhancing of existent resource-sharing policies through predictions. Through the execution of distinct well-known benchmarks across different many-core/multi-core architectures, we show that:
\begin{itemize}
    \item We equal -- and sometimes beat -- the performance of state of the art policies that prioritize performance.
    \item Our policies also equal and, in some scenarios, beat the energy efficiency of state of the art policies that prioritize energy efficiency.
    \item Enhancing resource-sharing techniques through predictions simultaneously improves performance and energy efficiency.
\end{itemize}

The remainder of this paper is structured as follows. In Section~\ref{section:background}, we discuss state of the art resource managing strategies in different parallel programming models. Next, in Section~\ref{section:improving-policies}, we give insight into our monitoring and prediction infrastructure and improved prediction-based policies. In Section~\ref{section:evaluation}, we present the evaluation of our proposals across different systems and various benchmarks. In Section~\ref{section:related-work} we go over related work and state of the art policies. Finally, in Section~\ref{section:conclusions-future-work} we give concluding remarks and comment on future work.
\section{Background}
\label{section:background}
In this paper, we study the enhancement of performance and energy efficiency through resource management policies for one of the most widespread parallelism strategies: tasking. In tasking, parallelism is specified through tasks, -- i.e., the basic unit work -- which are blocks of code that can be executed concurrently. The data flow of an application is specified through dependencies between tasks, which are annotated by users. OpenMP and OmpSs-2 are some programming models that can be used to exploit task-level parallelism. In OpenMP, users define parallelism through regions of code in which two or more threads may execute simultaneously. On the other hand, in OmpSs-2, there is an implicit parallel region that covers the whole application. This allows resource management to be more malleable since, at any point in the execution, the runtime system can idle or resume threads.

Regardless of the programming model, threads that are not doing useful computation at a given time -- e.g., while they are in a barrier -- must wait for a new workload. While waiting, threads behave differently depending on the underlying resource managing policies. Next, we describe conventional policies in the literature, along with their advantages and flaws.

\textbf{Active or Busy Policies:} In these, waiting threads are kept busy-waiting until work is available. Depending on the underlying runtime, this policy allows for an instant reaction to the creation of work. Nevertheless, it is a static policy that cannot adapt to workload changes. This exposes two main drawbacks in most OpenMP implementations. The first one is dealing with the contention caused by threads constantly polling shared data structures. In OmpSs-2, this problem is resolved through subscription locking techniques. However, energy efficiency -- the second drawback -- is ignored in the policies of both models, as threads consume processor cycles while busy-waiting.

\begin{figure}[b]
\centering
\includegraphics[width=\textwidth]{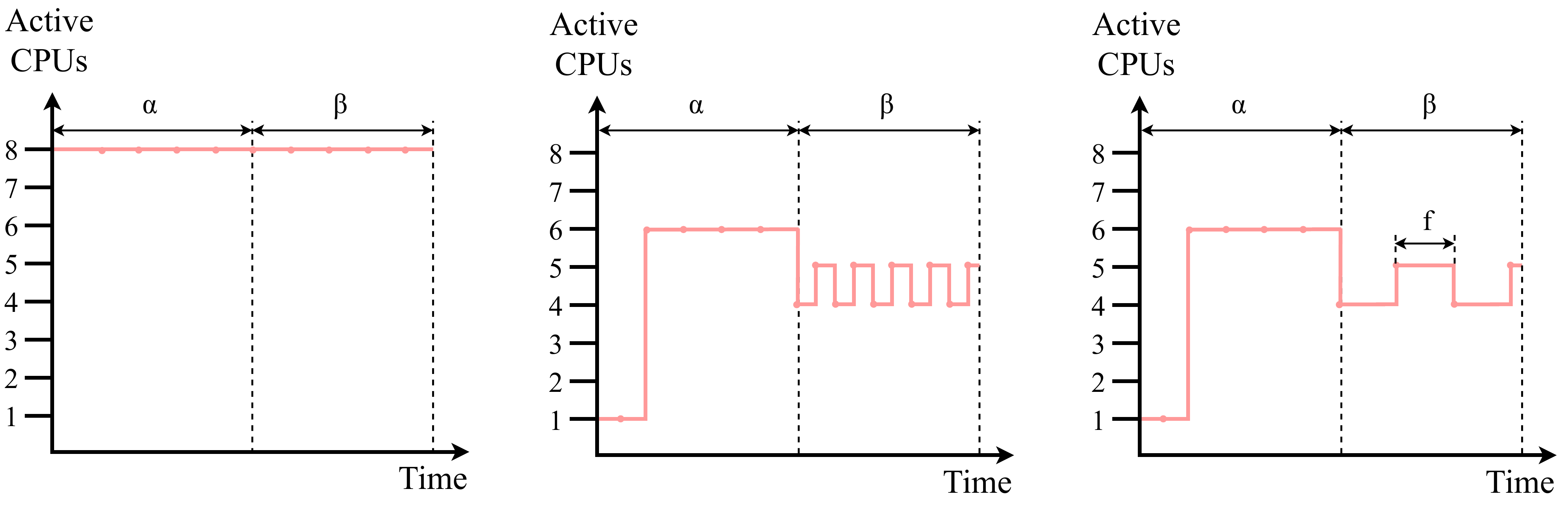}
\caption{Behavior of busy (left), idle (middle), and prediction policies (right)}
\label{figure:all_policy_behavior}
\end{figure}

Figure~\ref{figure:all_policy_behavior} exemplifies the number of active CPUs over time for a parallel region which has two different workload phases. The first phase ($\alpha$) has enough work for six CPUs, while the second one ($\beta$) has enough work, on average, for four and a half CPUs. In this scenario, in busy policies -- left part of the figure -- at all times there are eight threads actively polling for work.

\textbf{Passive or Idle Policies:} In these, waiting threads do not consume processor cycles. These policies are usually not reactive, as they are implemented by idling threads for a constant amount of time. This causes benefits for energy efficiency but may be adverse for performance. In OmpSs-2, as tasks are created, threads are resumed so they may poll once again. This allows for an instantaneous reaction to the addition of work, which makes it more reactive than OpenMP. Taking into account the previous example, the middle part of Figure~\ref{figure:all_policy_behavior} shows that, in these policies, threads are regularly being resumed and idled onto CPUs as the workload varies. Often, in fine-grained or irregular applications, this causes substantial amounts of overhead.

\textbf{Hybrid Policies:} To solve all the previously listed issues, OpenMP users can tune the rate at which waiting threads are idled and resumed. This enables users to find a balance between energy efficiency and performance. However, the chosen rate is a static value that cannot be changed at run-time. Therefore, this method cannot cope with variability in irregular applications, as these may need different rates throughout their executions.

\textbf{Resource Sharing:} OmpSs-2 offers an execution mode that integrates Dynamic Load Balancing (DLB). DLB~\cite{dlb09} is a tool that is transparent to users and enables runtimes or applications to share processing elements between each other. This sharing is implemented through the Lend When Idle (LeWI) mechanism. It showcases similarities when compared to the idle policy. When threads poll for tasks and receive none, the CPU onto which they are executing is shared -- instead of being idle. For this reason, depending on the application, this policy is excessively reactive and makes adverse decisions when lending/acquiring CPUs.
\section{Improving Resource Managing Policies}
\label{section:improving-policies}
As previously discussed, policies that do not look ahead are too naive to cope with the challenge of enhancing both energy efficiency and performance. Therefore, we advocate for policies that take into account workload predictions to make better decisions when handling processing elements. Next, we describe (\textit{i}) the necessary elements to create a monitoring and prediction infrastructure to equip runtime systems with the required information to create better policies, and (\textit{ii}) our approach towards finding a solution to the trade-off between energy efficiency and performance with prediction-based resource managing policies.

\subsection{Monitoring and Prediction Infrastructure}
\label{subsection:monitoring-infrastructure}
In order to tackle the drawbacks of current policies and the challenges exposed in Section~\ref{section:background}, we used a lightweight infrastructure capable of providing precise predictions with negligible overhead. Our infrastructure pinpoints critical changes in tasks, threads, or CPUs. Whenever possible, these changes are tracked outside the critical path of the runtime (i.e., synchronization points) so that the module is as lightweight as possible. Furthermore, to produce negligible overhead in fine-grained task scenarios, we combine the usage of atomic structures and the aggregation of metrics in a per-thread and per-task type basis.

\begin{figure}[t]
\centering
\includegraphics[width=0.7\textwidth]{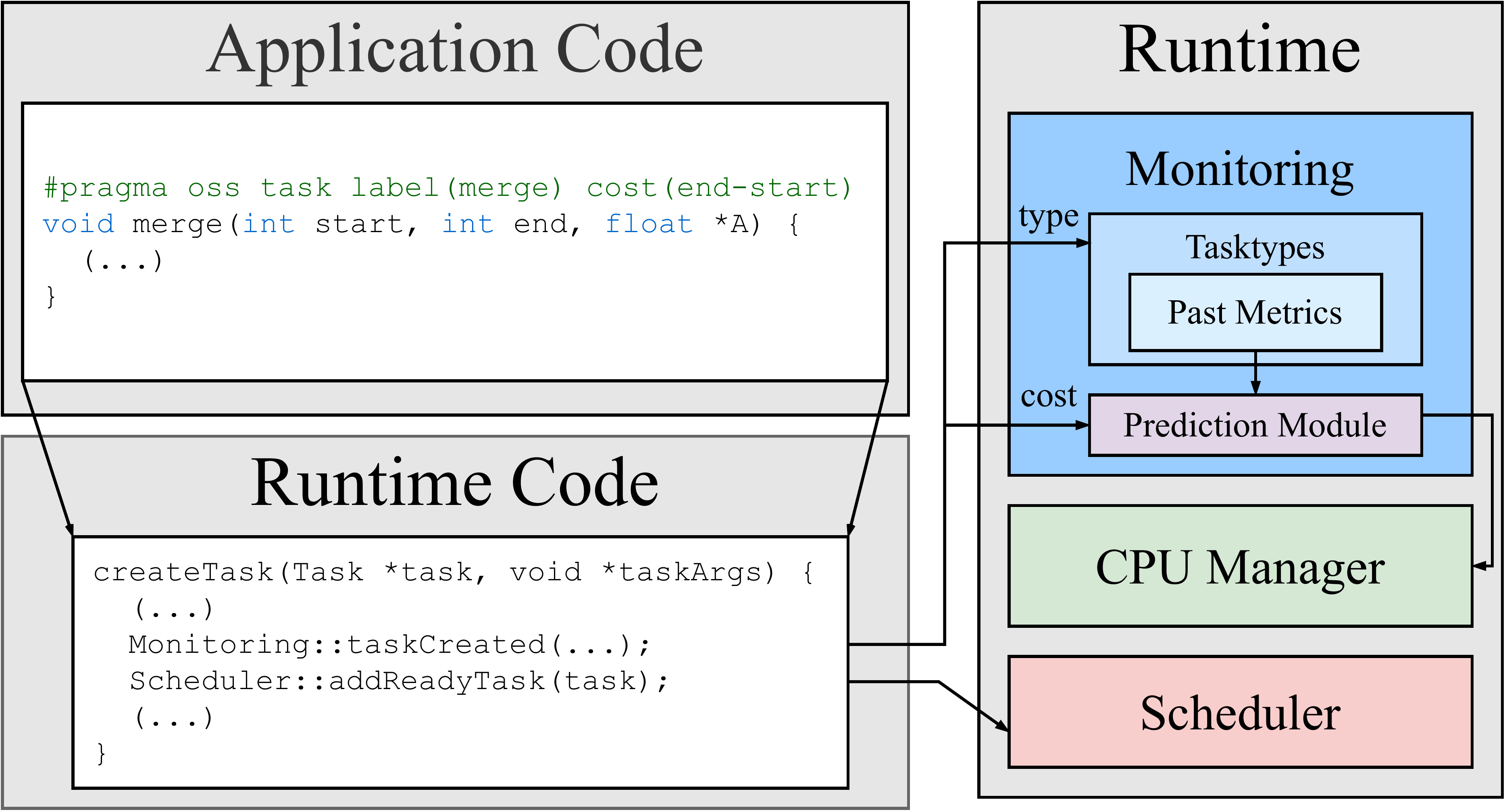}
\caption{A glimpse of the monitoring and prediction infrastructure}
\label{figure:monitoring_infrastructure}
\end{figure}

Another critical attribute is the precision of predictions. Averaging task execution times is not precise enough due to the variability discussed in previous sections. On top of that, two tasks of the same type may behave unexpectedly depending on their input size. For instance, one of the inputs may be too large to fit within the same cache hierarchy level. Thus, to solve this, we use the cost clause, already proposed in previous works~\cite{autofinal17}. This clause specifies, in a rough way, the computational weight of a task. Such information allows normalizing metrics in order to extrapolate predictions for any task of the same type. Furthermore, this clause is user-friendly and requires little effort, as its filler value should be well-known to application developers. Figure~\ref{figure:monitoring_infrastructure} generally exposes all the elements involved in the computation of predictions. Upon a task is created and placed in the scheduler, the monitoring module predicts its metrics using past information from similar tasks. Predictions are then accumulated and passed onto the prediction module, which aids the resource manager by predicting the number of resources to use for the current workload.

\begin{algorithm}[t]
\caption{Algorithm to predict the optimal CPU utilization ($\Delta$)}
\begin{multicols}{2}
\begin{algorithmic}[1]
\Statex $T_i$: Execution time of task $i$
\Statex $C_i$: Cost of task $i$
\Statex $f$: Prediction frequency
\Statex $N_{CPUs}$: Maximum number of CPUs
\Statex $W_{i_j}$: Workload for runtime status $i$ and tasktype $j$
\Statex $\alpha_j$: Normalized cost for tasktype $j$
\Statex $M_j$: Number of tasks of type $j$
\Ensure $0 < \Delta \leq N_{CPUs}$
\Statex 
\Statex
\Statex
\Statex
\Function{getCPUPrediction}{...}
\State $\gamma \leftarrow 0$
\State $j \leftarrow 0$
\While {$(\gamma < N_{CPUs}$)}
\State $\beta \leftarrow \frac{(W_{ready_j} + W_{execution_j}) * \alpha_j}{f}$
\State $\gamma \leftarrow \gamma + \beta$
\State $j \leftarrow j + 1$
\If{$i > N_{runtime\_status}$}
\State \textbf{\textit{break}}
\EndIf
\EndWhile
\State $\Delta \leftarrow min(\gamma, \sum_{j=0}^{tasktype_n} M_j)$
\EndFunction
\end{algorithmic}
\end{multicols}
\label{algorithm:prediction_algorithm}
\end{algorithm}

Algorithm~\ref{algorithm:prediction_algorithm} shows a pseudo-code that describes how resource utilization predictions are computed. As previously mentioned, timing metrics are aggregated on a per-task type basis. This allows at any given time to have a precise prediction of the available workload ($W_{i_j}$) for every runtime status ($i$) and every task type ($j$). With these and normalized information from the execution of past tasks of every task type ($\alpha_j$), we can precisely approximate the elapsed execution time of the available workload ($\beta$). Once a prediction rate is chosen ($f$), we can compute the optimal number of CPUs to utilize over that period ($\Delta$), which takes into account the number of available tasks as well as their expected execution time. This information is then passed to the resource manager so that the current number of CPUs can progressively be trimmed or increased to meet the prediction. Finally, to adapt to variability with haste, the normalized metrics are computed using a rolling window, which weights past metrics by their occurrence. The more recent these previous metrics are, the more weight they have towards the computation of their respective $\alpha$.

\subsection{Adaptive Prediction-Based Policies}
\label{subsection:adaptive_prediction_based_policies}
Throughout Section~\ref{section:background}, we describe the main flaws of current resource managing policies. To enhance these policies, we propose predicting the optimal number of CPUs at every point in time and at run-time. In other words, at a point in time $T_i$, we decide the number of CPUs to be used until $T_i + f$, where $f$ is the time interval until the next prediction is made.

As shown in Section~\ref{subsection:monitoring-infrastructure}, our resource managing predictions are based on task timing predictions. To compute the latter, we normalize task timing metrics using their cost values in order to obtain normalized or \textit{unitary costs} per task type. These unitary costs roughly represent the amount of time spent in the execution for each unit of cost of the task~\cite{autofinal17}. Then, we aggregate task costs per task type and runtime status separately. With these two metrics, at run-time, we compute the product of the accumulation of cost of all the task instances of a specific type by the respective unitary cost metric. Since these unitary values may vary over time, computing the product at run-time makes it susceptible to changes, which is precisely our goal. Furthermore, we average these unitary metrics using exponential moving averages. This allows them to be susceptible to variability and update as executions progress.

To compute the current amount of available workload in the system, we take into account ready and executing tasks. However, tasks in the executing status cannot account for their entire predicted time, as they may already be deep into their execution. To solve this, we aggregate task execution times through the parent-child link between tasks. When a task finishes, its execution time is subtracted from the parent's task predicted time, if it is available.

\begin{algorithm}[t]
\caption{Pseudo-code of the behavior of threads within the CPU manager}
{\scriptsize
\begin{algorithmic}[1]
\Statex $\Delta$: An atomic variable that holds the predicted optimal number of CPUs
\Statex $\delta$: The current number of active CPUs
\Statex $a$: The action that triggered the call (polling or adding tasks)
\Function{executePolicy}{thread, a}\Comment{$\delta$ is updated in a thread-safe manner}
\If{$a == POLL$}
\If{$queue == \emptyset$}
\If{$\delta > \Delta$}
\State $\delta \leftarrow \delta - 1$
\State idle(thread)
\State cpu $\leftarrow$ getCPU(thread) 
\State releaseCPU(cpu)
\EndIf
\EndIf
\Else\ [$a == ADD$]
\If{$\delta < \Delta$}
\State idleThread $\leftarrow$ getIdleThread()
\State idleCPU $\leftarrow$ acquireCPU()
\If{idleCPU $\neq \emptyset$}
\State $\delta \leftarrow \delta + 1$
\State resume(idleCPU, idleThread)
\EndIf
\EndIf
\EndIf
\EndFunction
\end{algorithmic}
}
\label{algorithm:pseudocode_cpumanager}
\end{algorithm}

In Algorithm~\ref{algorithm:pseudocode_cpumanager}, we show a pseudo-code of how our CPU manager uses these predictions. Rather than forcing the runtime to comply with the predicted number of CPUs ($\Delta$), we save this value in an atomic variable. Then, when threads poll for tasks and none exist, if this value marks that the current number of active CPUs must be decreased, the thread idles until further notice, so that it does not consume CPU cycles. Reversely, when tasks are added into the scheduler and this value marks that more CPUs are required, idle threads are resumed to execute these newly created tasks.

The main benefits of our policy are twofold. If we compare our prediction policy to the idle or passive policies, a common feature is that they are both highly reactive to changes in the available workload. However, predictions occur at a specific rate. This allows our policy to avoid the overhead of continuously waking and idling threads in fine-grained or irregular applications. This benefit can also be seen as a middle ground between idle and busy policies. Taking into account the example introduced in Section~\ref{section:background}, Figure~\ref{figure:all_policy_behavior} shows the behavior of our prediction policy (right part). The rate at which predictions are inferred avoids multiple idling and resuming operations which, in the long run, adds up to avoid substantial overhead.

Another primary benefit of our policy is the adaptiveness to the granularity of tasks. Managing resources by only considering the number of ready tasks is enough in some scenarios. Nonetheless, for applications with fine-grained tasks, it would end up utilizing an excessive amount of CPUs for their workload. With the prediction policy this is resolved, as it takes into account the predicted granularity of tasks.

\subsection{Prediction-Based Sharing of Resources}
\label{subsection:prediction_based_sharing_resources}
Section~\ref{section:background} briefly introduces the DLB integration of OmpSs-2. This mode of execution, as previously mentioned, resembles the idle policy. As it is as reactive, it may produce huge amounts of calls to the DLB library when lending or acquiring CPUs. Such calls do not come for free; they introduce non-negligible overhead.

To fix this flaw, we propose to modify the mechanism within OmpSs-2 to avoid making eager decisions. Our idea follows the same concept as the one adopted in the prediction policy. Instead of letting threads decide when CPUs are lent or acquired, we offload such decisions to an external prediction heuristic. Similarly, this heuristic predicts the amount of workload currently available in the system. Nonetheless, it is slightly modified to allow a superior number of CPUs, as DLB may provide more CPUs than the ones currently available to the runtime. When a thread polls for tasks and receives none, it will use the heuristic to decide whether its CPU must be lent. Simultaneously, as soon as predictions are inferred, the heuristic makes a single call to DLB in order to acquire as many CPUs as required. Therefore, threads do not require to do it progressively.
\section{Experimental Setup}
\label{section:experimental_setup}

The experiments we performed were run on Intel Xeon and KNL multi-core systems, as shown in Table~\ref{table:systems}. In the same table, we also show the compilers used in each system. We present all results as the arithmetic mean of five runs for all metrics. To measure the energy efficiency, we consider the energy-delay product (EDP), which correlates both performance and energy consumption in only one value. To retrieve energy consumption metrics, we used the Intel Running Average Power Limit library~\cite{intelrapl}. The evaluation is partitioned into two phases. The first phase targets the measuring of overhead of our strategies and a comparison between the policies in two versions of OmpSs-2 and different OpenMP implementations. The second targets the evaluation of our prediction-based strategy for resource sharing using DLB.

\begin{table}[b]
\centering
{\scriptsize
\caption{Architectures used in our experimental setup}
\resizebox{0.7\textwidth}{!}{
\begin{tabular}{|l|l|l|l|}
\cline{1-1} \cline{3-4}
\textbf{Name}    &  & \textbf{MN4}              & \textbf{KNL}    \\ \cline{1-1}
\cline{3-4}
Processor        &  & Intel Xeon Platinum 8160  & Intel Xeon Phi CPU 7230   \\
Architecture     &  & Skylake                   & Knights Landing           \\
Frequency        &  & 2.10GHz                   & 1.30GHz                   \\
\# of Sockets    &  & 2                         & 1                         \\
\# of Cores      &  & 48 (24 x 2)               & 64                        \\
Memory           &  & 96 GB                     & 96 GB                     \\
OS               &  &SUSE 12 SP2                & SUSE 12 SP2               \\
Intel Compiler   &  & 19.1.0.166                & 19.1.0.166                \\
GNU Compiler     &  & 9.2.0                     & 9.2.0                     \\ \cline{1-1} \cline{3-4} 
\end{tabular}
}
\label{table:systems}
}
\end{table}

In our experiments, we used the \textit{Cholesky Factorization} benchmark and the \textit{High Performance Computing Conjugate Gradients}\footnote[1]{HPCCG is implemented using multidependences, available in OpenMP 5.0~\cite{openmp}. As the Intel 2020.0 compiler does not support them, HPCCG-\textit{IOMP} results are missing.} (HPCCG) mini-application. The former decomposes a matrix into the product of a lower triangular matrix and its conjugate transpose. The latter is based on the CG benchmark for a 3D chimney domain. They are both highly scalable benchmarks that present varying compute-intensive workloads. Furthermore, to test irregularity in applications, we used two versions of Cholesky; one that produces coarse-grained tasks, and another that creates an excessive amount of fine-grained tasks. Similarly, we also covered both granularity scenarios for the \textit{MultiSAXPY} benchmark, which performs the SAXPY level one operation from the Basic Linear Algebra Subprograms package~\cite{blas79}. Finally, to test our policies in memory-bound benchmarks, we used \textit{Gauss-Seidel} and \textit{STREAM}. The former is a solver that simulates the distribution of heat over time, and the latter is a benchmark that measures memory transfer rates in MB/s. While Gauss-Seidel could be highly parallel, to have a fair comparison against OpenMP, in the OmpSs-2 implementation we include a barrier after each time step. This produces load imbalance but, simultaneously, makes it an ideal candidate to be combined with STREAM, which is highly parallel and balanced.

\section{Evaluation}
\label{section:evaluation}

Even though assessing the accuracy of our predictions was done in previous works, in Table~\ref{tab:accuracy-table} we include results of the accuracy of task timing predictions for all benchmarks and machines. In this table, we showcase the number of task instances used to compute the accuracy results and the average accuracy of all predictions. These predictions are then used towards calculating the optimal number of CPUs to use, as shown in Algorithm~\ref{algorithm:prediction_algorithm}. The \textbf{(F)} and \textbf{(C)} shown next to benchmark names indicate whether the results are for the fine-grained scenario or the coarse-grained scenario, respectively. Due to the low number of task instances in coarse-grained Cholesky, CPU utilization predictions are based only on the number of available tasks, which is the go-to approach when task timing predictions are not available. Throughout the whole evaluation we used the same prediction rate -- \texttt{\textit{f}} in Algorithm~\ref{algorithm:prediction_algorithm} -- of 50 $\mu$s.

\begin{table}[b]
\caption{Average prediction accuracy of each benchmark and architecture}
\resizebox{\textwidth}{!}{%
\begin{tabular}{llcccccc}
\cline{3-8}
\textbf{}                                      & \multicolumn{1}{l|}{} & \multicolumn{6}{c|}{\textbf{MN4}}                                                                                                                                                                                                                                               \\ \cline{1-1} \cline{3-8} 
\multicolumn{1}{|l|}{\textbf{Benchmark}}       & \multicolumn{1}{l|}{} & \multicolumn{1}{l|}{\textbf{Cholesky (F)}} & \multicolumn{1}{l|}{\textbf{Cholesky (C)}} & \multicolumn{1}{l|}{\textbf{HPCCG}} & \multicolumn{1}{l|}{\textbf{Gauss-Seidel}} & \multicolumn{1}{l|}{\textbf{Multisaxpy (F)}} & \multicolumn{1}{l|}{\textbf{Multisaxpy (C)}} \\ \cline{1-1} \cline{3-8} 
\multicolumn{1}{|l|}{\textbf{\# of Instances}} & \multicolumn{1}{l|}{} & \multicolumn{1}{c|}{3*10$^6$}  & \multicolumn{1}{c|}{600}                        & \multicolumn{1}{c|}{15000}         & \multicolumn{1}{c|}{25600}         & \multicolumn{1}{c|}{1*10$^5$}    & \multicolumn{1}{c|}{20000}                        \\ \cline{1-1} \cline{3-8} 
\multicolumn{1}{|l|}{\textbf{AVG Accuracy}}    & \multicolumn{1}{l|}{} & \multicolumn{1}{c|}{88.25\%}                  & \multicolumn{1}{c|}{NA}                         & \multicolumn{1}{c|}{78.45\%}       & \multicolumn{1}{c|}{99.91\%}       & \multicolumn{1}{c|}{70.63\%}                    & \multicolumn{1}{c|}{79.49\%}                      \\ \cline{1-1} \cline{3-8} 
                                               &                       & \multicolumn{1}{l}{}                          & \multicolumn{1}{l}{}                            & \multicolumn{1}{l}{}               & \multicolumn{1}{l}{}               & \multicolumn{1}{l}{}                            & \multicolumn{1}{l}{}                              \\ \cline{3-8} 
\multicolumn{1}{c}{\textbf{}}                  & \multicolumn{1}{l|}{} & \multicolumn{6}{c|}{\textbf{KNL}}                                                                                                                                                                                                                                               \\ \cline{1-1} \cline{3-8} 
\multicolumn{1}{|l|}{\textbf{\# of Instances}} & \multicolumn{1}{l|}{} & \multicolumn{1}{c|}{3*10$^6$}  & \multicolumn{1}{c|}{600}                        & \multicolumn{1}{c|}{15000}         & \multicolumn{1}{c|}{25600}         & \multicolumn{1}{c|}{1*10$^5$}    & \multicolumn{1}{c|}{20000}                        \\ \cline{1-1} \cline{3-8} 
\multicolumn{1}{|l|}{\textbf{AVG Accuracy}}    & \multicolumn{1}{l|}{} & \multicolumn{1}{c|}{92.65\%}                  & \multicolumn{1}{c|}{NA}                         & \multicolumn{1}{c|}{75.32\%}       & \multicolumn{1}{c|}{99.81\%}       & \multicolumn{1}{c|}{76.83\%}                    & \multicolumn{1}{c|}{86.12\%}                      \\ \cline{1-1} \cline{3-8} 
\end{tabular}%
}
\label{tab:accuracy-table}
\end{table}

To measure the overhead of our monitoring infrastructure, we ran all the previously mentioned benchmarks with varying task granularities. We compared OmpSs-2’s current busy policy against a modified version of the busy policy that monitors metrics and infers predictions, but uses neither. We observed that for extreme situations with millions of fine-grained tasks, our infrastructure adds, in the worst case, a maximum overhead of \textbf{3\%} to the execution time. We believe these overheads are negligible in comparison to the benefits we obtain.

\begin{figure*}[!b]
    \centering
    \includegraphics[width=\textwidth]{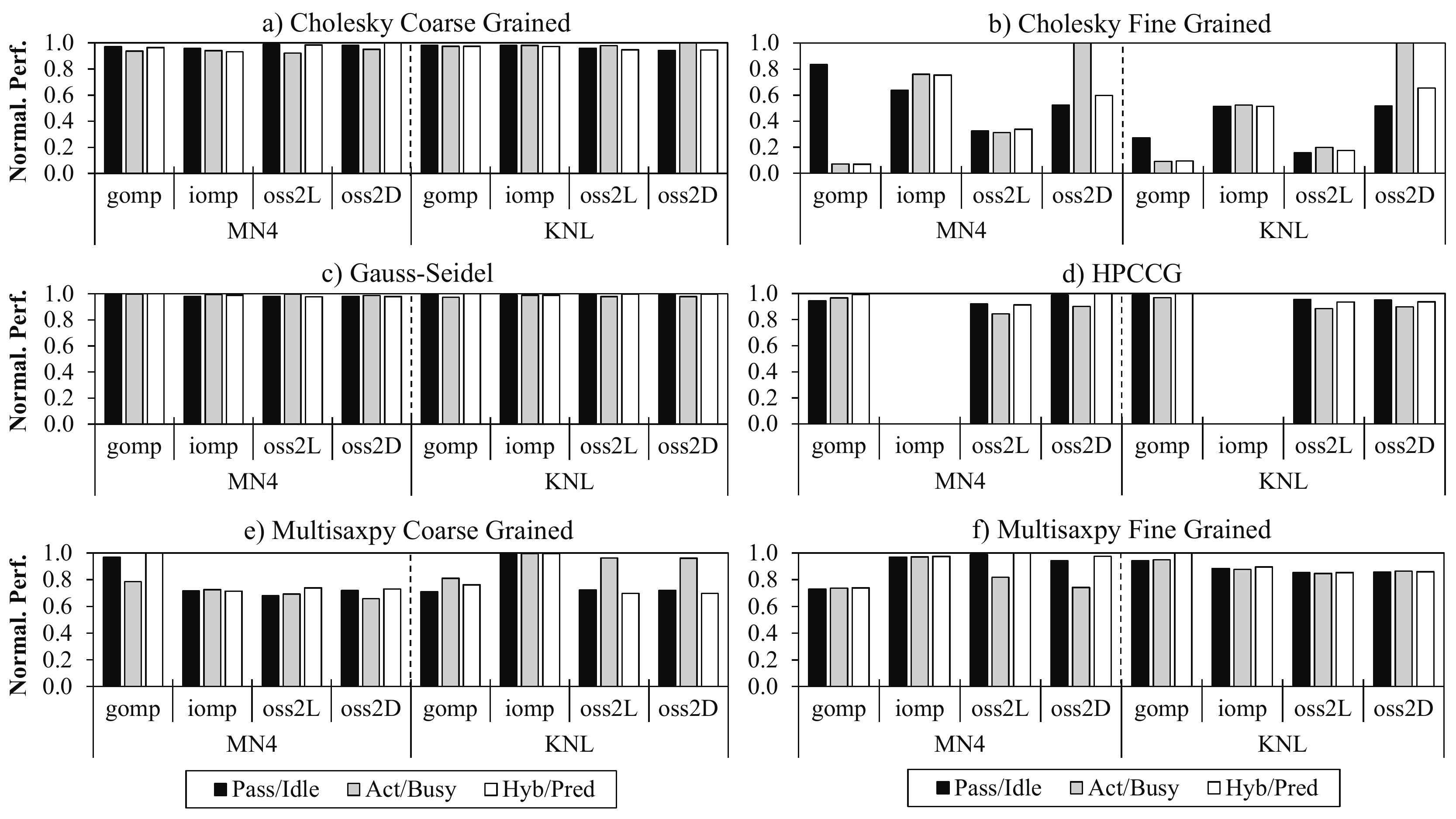}
    \caption{Normalized performance w.r.t. the best scenario on each application}
    \label{fig:time_results}
\end{figure*}

Our evaluation comprises four different implementations: GCC OpenMP~\cite{Novillo06openmpand} (\textbf{gomp}), Intel OpenMP~\cite{intelopenmp} (\textbf{iomp}), OmpSs-2 using its linear regions dependency system (\textbf{oss2L}), and OmpSs-2 using its improved discrete dependency system (\textbf{oss2D}). For the OpenMP implementations, we evaluate all their available thread-waiting policies: \textbf{active}, \textbf{passive}, and a \textbf{hybrid} between both. For the OmpSs-2 counterparts, we evaluate their current resource managing policies, \textbf{busy} and \textbf{idle}, and our \textbf{prediction} policy. Due to the similarities in their concepts, we group the comparison as follows: Active/Busy, Passive/Idle, and Hybrid/Prediction. In all figures, from left to right, we show the results of gomp, iomp, oss2L, and oss2D.

Figure~\ref{fig:time_results} showcases the normalized performance of all benchmarks, architectures, and between all policies. For Cholesky's coarse-grained scenario, Gauss-Seidel, and HPCCG, the performance obtained using the prediction policy in both OmpSs-2 versions either equals or surpasses the performance of all other policies in MN4. In fine-grained Multisaxpy, comparing all the OmpSs-2 policies, our policy yields either similar performance (in KNL) or surpasses other policies (in MN4). Nonetheless, in the coarse-grained scenario in KNL, busy yields better performance than prediction. We attribute this to the precision in predictions, as shown in Table~\ref{tab:accuracy-table}. This accuracy could be enhanced by taking into account other metrics -- as it is a memory-bound benchmark. Last but not least, in the fine-grained Cholesky scenario, the prediction policy yields similar performance when compared to OmpSs-2's linear version. However, in the discrete version, its performance remains between the busy and idle policies, being busy the most performant. This difference between versions led us to find out that the monitoring infrastructure adds slightly more overhead in OmpSs-2's discrete version, as contention is minimal and the overhead shifts to other runtime modules.

\begin{figure*}[!b]
    \centering
    \includegraphics[width=\textwidth]{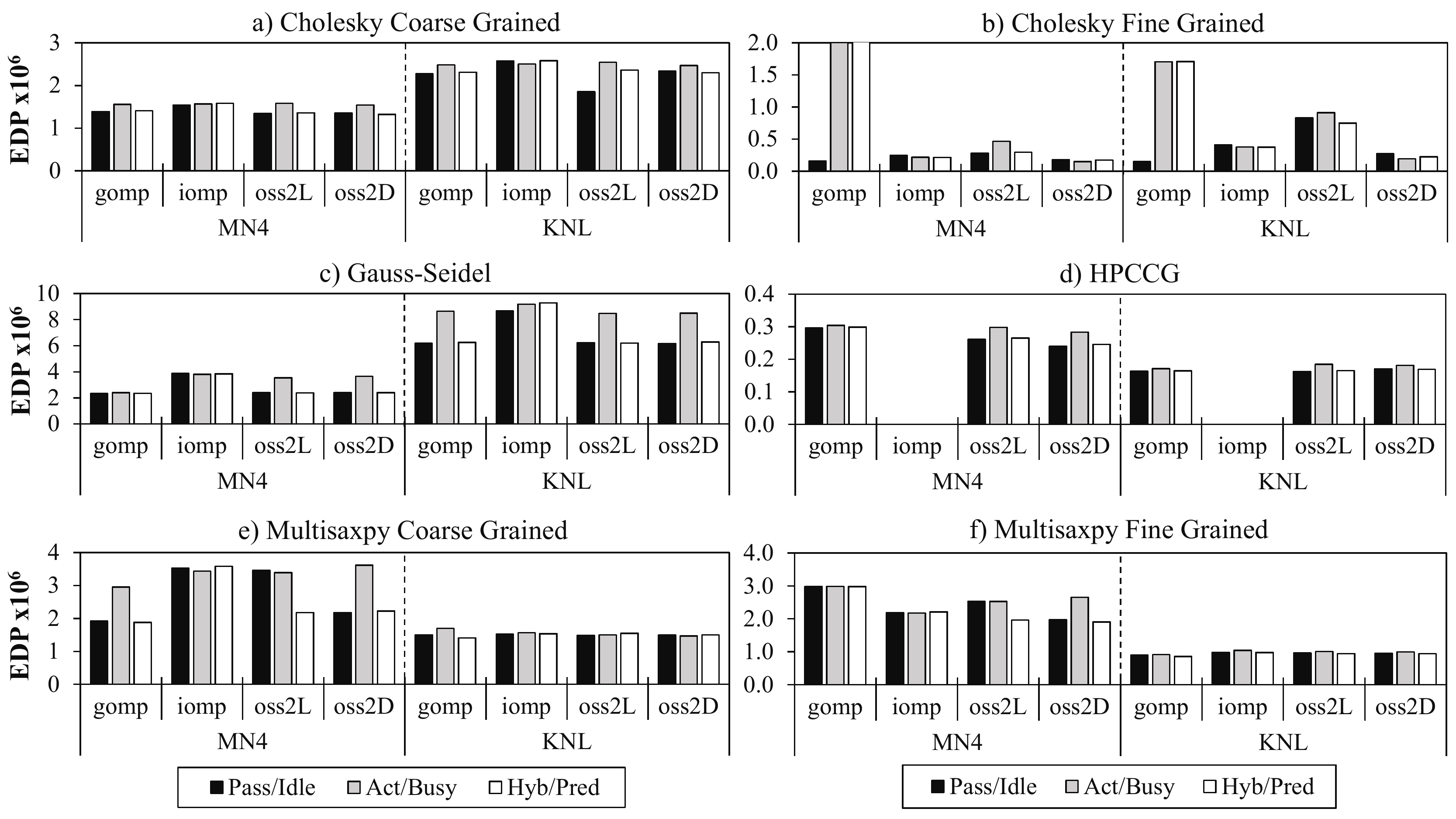}
    \caption{EDP results (raw data) for each application and policy}
    \label{fig:edp_results}
\end{figure*}

Figure~\ref{fig:edp_results}, shows the comparison of EDP between all policies and architectures. Thus, in these plots, lower values are better. For coarse-grained Cholesky, prediction policies obtain better results than any other policies in MN4. In KNL, the only configuration that beats the prediction policy of OmpSs-2 discrete is OmpSs-2 linear's idle policy. As for the fine-grained scenario, OmpSs-2 discrete's prediction policy yields less EDP than any other policy for both architectures except when compared to GOMP's passive policy in KNL, as their results are similar. In both Gauss-Seidel and HPCCG, prediction policies beat any other policy in any implementation and architecture. Finally, for the coarse-grained Multisaxpy scenario, EDP results in KNL are very similar across policies and implementations. However, in MN4, prediction policies achieve considerably lower EDP than any other policy except when compared to GOMP's hybrid policy, which obtains similar results. Both fine-grained and coarse-grained scenarios present similarities. However, as predictions benefit from fine-grained and irregular applications, in MN4 prediction policies beat any other policy in EDP.

To further evaluate our predictions, we created a prediction-based policy for the DLB execution mode of OmpSs-2. We chose to run the Gauss-Seidel simulation along with the STREAM benchmark in MN4, as they vary in features and, thus, combine perfectly when executed concurrently. For the former, we used an input size that generates slightly coarse-grained tasks, while in the latter, we chose an input size that generates fine-grained tasks. Thus, STREAM benefits from the lack of workload of Gauss-Seidel after each time-step. In Table~\ref{table:dlb-results}, we show the average results of several executions with multiple configurations. We executed both applications concurrently, each in a single NUMA node (half the number of processors of the whole node) for the \textit{Concurrent} configuration. To take into account any possible noise between shared resources -- e.g., cache pollution or bandwidth thresholds -- we also executed each application in a single NUMA node on its own, which is shown as the \textit{Single} configuration. Then, we executed using DLB to share cores between applications in three configurations: \textit{Concurrent + DLB LeWI} is the default policy. \textit{Concurrent + DLB Hybrid} is a modified version of the DLB integration that only shares CPUs after several failed attempts of polling tasks from the scheduler -- hence the similar name to OpenMP's hybrid policy. For our experiments we chose 100 as the number of attempts before a CPU is shared. Finally, \textit{Concurrent + DLB Prediction} shows the results of the DLB execution mode enhanced with our predictions.

In the \textit{LeWI} policy, STREAM can benefit from the lack of workload of Gauss-Seidel, thus reducing its execution time. Nevertheless, as this policy is extremely reactive, the combination of the number of calls to DLB is around 4 million in executions of 100 and 75 seconds, respectively. These calls add non-negligible overhead. On top of that, since Gauss-Seidel lends CPUs for short amounts of time in which neither applications can benefit, its execution time increases. To try to tackle this flaw in fine-grained scenarios, we let threads spin for a while before lending their CPU in the \textit{Hybrid} version. Nevertheless, as shown, the number of calls and execution time are similar to \textit{LeWi's}. By spinning before lending, the runtime is stressed with more contention, thus leading to similar execution times, EDP, and number of DLB calls. Finally, when enhancing the \textit{LeWI} policy with predictions (\textit{DLB Prediction}), the results are promising. As shown, the number of DLB calls is greatly reduced -- \textbf{4} times fewer calls. Simultaneously, better decisions are taken both when lending and acquiring CPUs. This leads to a \textbf{1.4x} speedup for STREAM, similar execution times for Gauss-Seidel, and a considerable reduction in EDP in STREAM as well. Furthermore, when comparing EDP metrics between policies that use DLB and the \textit{Single} policy, it is noticeable that results are worse for the DLB counterparts. Since the \textit{Single} policy idles CPUs when they are not used, EDP is better than in DLB policies where CPUs are never idled. Hence, if energy metrics are the primary target, the \textit{Prediction} policy in the non-DLB scenario would be preferable.

\begin{table}[t]
\caption{Comparison of metrics between OmpSs-2 + DLB policies}
\label{table:dlb-results}
\centering
\setlength{\tabcolsep}{0.2em}
{\scriptsize
\begin{tabular}{|l|l|r|r|r|r|r|r|r|}
\cline{1-7} 
\multicolumn{1}{|c|}{\multirow{2}{*}{\textbf{Config.}}} &  \multicolumn{2}{c|}{\textbf{Time (s)}} & \multicolumn{2}{c|}{\textbf{EDP}} &  \multicolumn{2}{c|}{\textbf{\# DLB Calls}} \\ \cline{2-7} 
\multicolumn{1}{|c|}{} &  \multicolumn{1}{c|}{\textit{Gauss-Seidel}} & \multicolumn{1}{c|}{\textit{Stream}} & \multicolumn{1}{c|}{\textit{Gauss-Seidel}} & \multicolumn{1}{c|}{\textit{Stream}} & \multicolumn{1}{c|}{\textit{Gauss-Seidel}} & \multicolumn{1}{c|}{\textit{Stream}} \\ \cline{1-7} 
\textit{Single}  & 87.4 & 78.3 & 1097330
 & 1223111  & -- & -- \\ \cline{1-7}
\textit{Concurrent} & 87.4 & 78.3 & 1705625 & 1434911 &  -- & --  \\ \cline{1-7}
\textit{Concurrent + DLB LeWI} &  100.9 & 73.3 & 2318388 & 1427072 & 768078 & 3392692 \\ \cline{1-7}
\textit{Concurrent + DLB Hyb.} &  101.5 & 74.3 & 2318244 & 1471916 & 890924 & 3169858 \\ \cline{1-7}
\textit{Concurrent + DLB Pred.} &  89.6 & 55.8 & 1849275 & 870269 &  69504 & 842634 \\ \cline{1-7}
\end{tabular}
}
\end{table}

\begin{figure*}[!b]
    \centering
    \minipage{0.50\textwidth}
        \includegraphics[width=\linewidth]{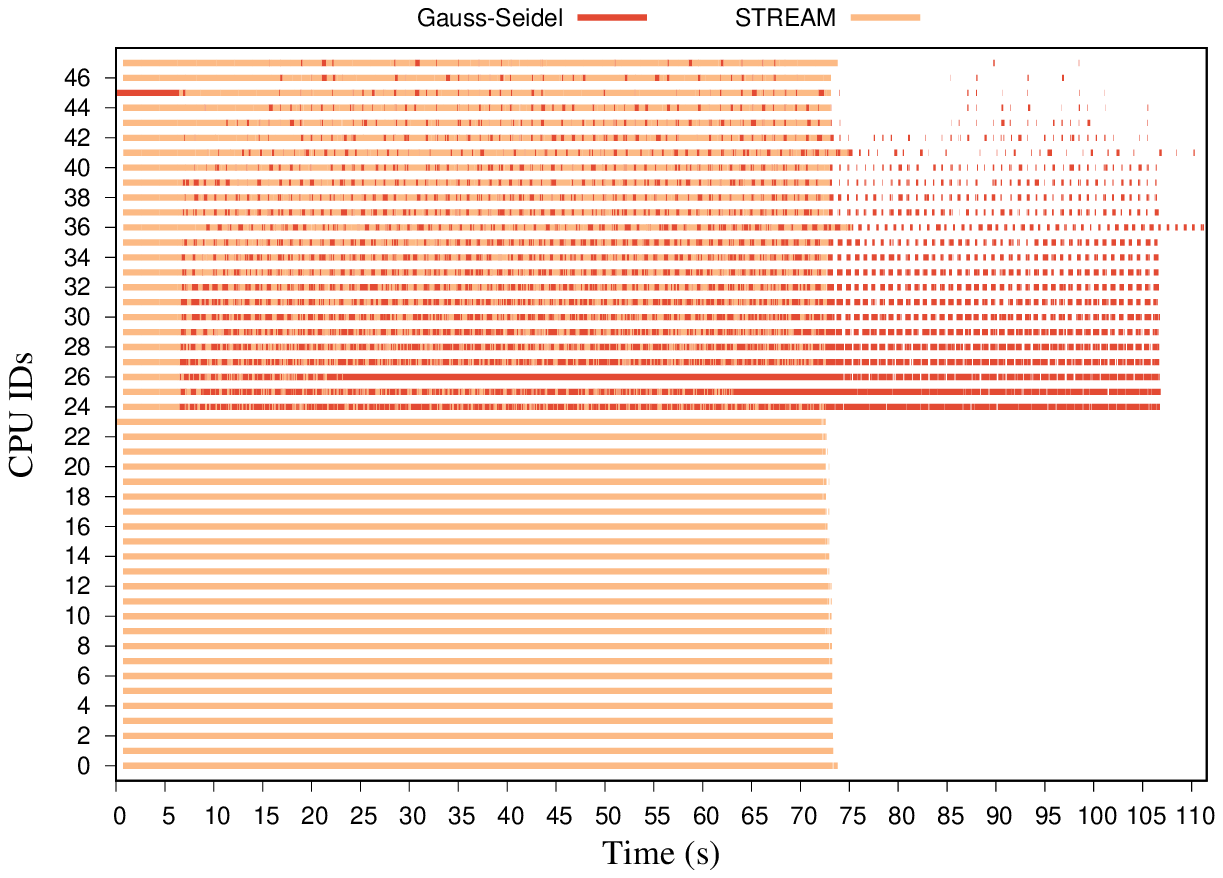}
    \endminipage
    \minipage{0.50\textwidth}
        \includegraphics[width=\linewidth]{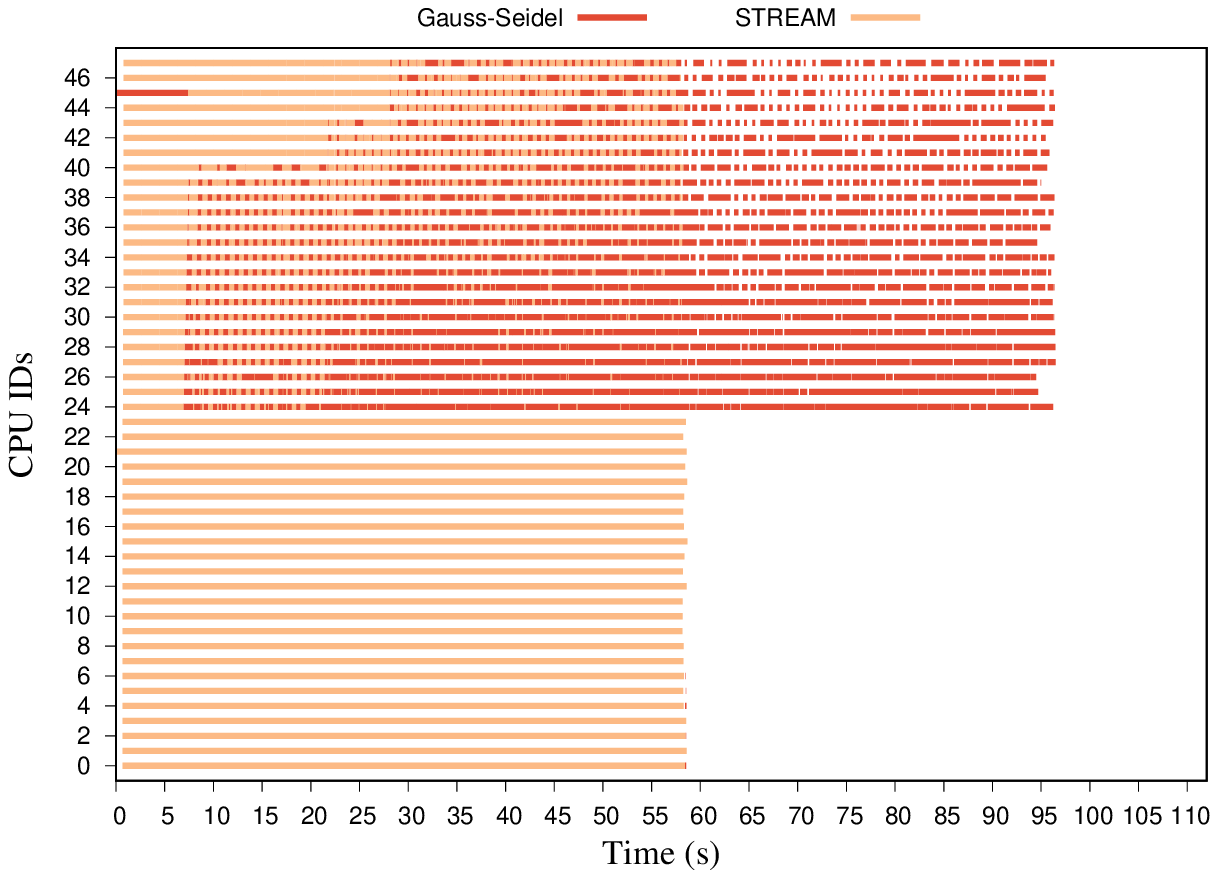}
    \endminipage
    \caption{Execution traces of hybrid (left) and prediction (right) DLB policies}
    \label{fig:prediction-trace}
\end{figure*}

To visualize how prediction-based policies improve resource sharing, we add the execution traces of the previous scenario for both the \textit{DLB + Hybrid} policy (left) and the \textit{DLB + Prediction} policy (right) in Figure~\ref{fig:prediction-trace}. To shorten execution traces, the execution of these benchmarks was slightly different in granularity of tasks when compared to the results shown in Table~\ref{table:dlb-results}, hence the slight difference in execution time. The red series corresponds to the execution of Gauss-Seidel, while the light-orange series corresponds to STREAM. In the prediction policy, as shown, the granularity of sharing in resources is coarser. CPUs are not lent unless they will not be used for a certain amount of time, and they will not be acquired unless they truly are required. Reversely, in the hybrid policy, there are still flawed decisions when lending or acquiring CPUs. As shown, there is much sharing that could be removed to avoid both the delays in Gauss-Seidel and the overhead of lending and immediately after re-acquiring CPUs.
\section{Related Work}
\label{section:related-work}


\textbf{Resource Management:} Techniques aiming to improve performance through resource management have been thoroughly studied. Barekas et al.~\cite{barekas2003multiprogramming} and Callisto~\cite{harris2014callisto} advocate for inter-process sharing of resources in their proposals. The former presents a resource manager and a runtime system which, respectively, distribute hardware resources to OpenMP applications and adapt their degree of parallelism. Although it is capable of providing better performance than commercial implementations of OpenMP, their approach offers no policies to improve performance between parallel regions. Callisto is a resource management layer for parallel runtime systems that coordinates the execution of parallel applications. It consists of (\textit{i}) a dynamic scheduler that defines which jobs can execute in parallel; and (\textit{ii}) a low-level API to manage synchronization points. However, it assumes that parallel sections of jobs are CPU-bound and that runtime systems need to be adapted to use Callisto.
Eichenberger et al.~\cite{eichenberger2012design} propose a model to control thread affinity for OpenMP applications. However, their work does not present any advances regarding the optimization of resource management policies.

Other proposals~\cite{boguslavsky1994optimal,yan2016proposal} have focused on our primary target, optimizing resource policies to improve performance or maintain it while improving energy efficiency. In this line, Boguslavsky et al.~\cite{boguslavsky1994optimal} investigate different strategies to determine for how long processes should spin before blocking. Even though their results are promising, such static values cannot cope with irregular applications that may need different blocking rates throughout their executions. To deal with oversubscription in OpenMP applications, Yan et al.~\cite{yan2016proposal} define five policies: spin\_busy, spin\_pause, spin\_yield, suspend, and terminate. However, in OpenMP, such policies cannot change within parallel regions. Thus, this approach flaws similarly. On the other hand, our approach is capable of dealing with such situations, as our policies can adapt at any point in the execution.


\textbf{Thread Malleability within Parallel Regions:} A number of studies, such as Thread Reinforcer (TR)~\cite{Pusukuri-Reinforcer}, Feedback-Driven Threading (FDT)~\cite{Suleman-FDT}, and ACTOR~\cite{actor}, investigate on optimizing either performance or energy by tuning the number of threads in parallel regions. TR~\cite{Pusukuri-Reinforcer} is a framework in which applications are executed multiple times with varying numbers of threads. FDT~\cite{Suleman-FDT} adapts the number of threads by considering contention in locks and memory bandwidth. ACTOR~\cite{actor} is a system that aims to improve the energy efficiency of parallel applications. In it, artificial neural networks are used to predict the number of threads to execute each parallel region. These previous approaches require either warm-up executions or techniques that may introduce substantial amounts of overhead when done at runtime.
Several other studies target solutions at run-time. LIMO~\cite{Chadha:LIMO} is a system that monitors applications and adapts the execution accordingly. Parcae~\cite{parcae} is a framework that creates multiple parallel transforms of sequential programs and, at run-time, determines the degree of TLP exploitation. Similarly, ParallelismDial \cite{Sridharan:holistic} is a model that automatically regulates the number of threads per region. Nonetheless, some of these approaches tune applications specifically for input sets and architectures. Others require OS support to intercept blocked threads to change their policies.


\textbf{Energy Efficiency:} Improving energy efficiency through resource management policies has been investigated as well, in studies such as OpenMPE~\cite{Alessi2015}, Benedict et al. \cite{7284455}, and LAANT~\cite{laant}. In the former, an OpenMP extension designed to improve energy management is proposed. In~\cite{7284455}, the authors propose an energy prediction mechanism for OpenMP applications using a Random Forest Modeling approach. LAANT~\cite{laant} is a library that aims at optimizing the EDP metric. The study conducted in Porterfield et al.~\cite{porterfield-2013} similarly proposes a system to automatically adjust the number of threads based on on-line measurements of system resource usage. These works are based on adjusting the number of threads of OpenMP applications in parallel-regions or the whole application. Thus, similarly to our previous explanation, they lack adaptiveness when it comes to irregular applications.
Li et al.~\cite{ipdps2010} propose a library to reduce energy consumption for hybrid MPI/OpenMP applications. Even though their aim is out of our scope, they use prediction models to enhance energy efficiency with negligible or no loss of performance. Finally, Shafik et al.~\cite{Shafik:2015a} propose an adaptive energy minimization model for OpenMP programs using annotations. These annotations require execution time estimations, which leads us to believe warm-up executions are needed to provide the library with such metrics.
\section{Concluding Remarks \& Future Work}
\label{section:conclusions-future-work}
In this paper, we presented resource management policies based on predictions that simultaneously optimize performance and energy efficiency. More specifically, we showcase (\textit{i}) a prediction-based CPU managing policy that maintains performance while improving energy efficiency, and (\textit{ii}) a prediction-based resource sharing mechanism which enhances both performance and energy efficiency when compared to its predecessor. We exemplify our proposal in OmpSs-2, although our approach can be applied to other parallel programming models based on tasks or fork-join.

While our prediction-based policies are capable of making better decisions than well-known policies from state of the art, we left a few aspects out of the scope which we will target in future work. Firstly, our policies could benefit from taking prediction error into account. Thus, when detecting anomalies, our infrastructure would be able to swap between CPU managing policies at run-time. We also believe that the rate at which predictions are inferred may be improved with a combined approach that triggers our mechanism when a certain number of events happen -- e.g., the creation or finalization of a number of tasks. Finally, we also plan to enable an on-line task characterization so that both our policies and predictions take more than one metric into account.
\section{Acknowledgments}

This project is supported by the European Union's Horizon 2020 research and innovation programme under the grant agreement No 754304 (DEEP-EST), the Ministry of Economy of Spain through the Severo Ochoa Center of Excellence Program (SEV-2015-0493), by the Spanish Ministry of Science and Innovation (contract TIN2015-65316-P) and by the Generalitat de Catalunya (2017-SGR-1481). This work was also supported by Project HPC-EUROPA3 (INFRAIA-2016-1-730897), with the support of the EC Research Innovation Action under the H2020 Programme.

\bibliographystyle{splncs04} 
\bibliography{bib/references}

\end{document}